\documentclass[aps,prl,twocolumn,
amsmath,showpacs]{revtex4}
\usepackage{amssymb,graphics,graphicx, epsfig}
\newcommand{\newc}{\newcommand}
\newc{\beqa}{\begin{eqnarray}}
\newc{\eeqa}{\end{eqnarray}}
\newc{\beq}{\begin{equation}}
\newc{\eeq}{\end{equation}}
\newc{\nonr}{\nonumber}
\newc{\ra}{\rightarrow}
\newc{\tri}{\triangle}
\newc{\PD}{\partial}
\newc{\lag}{{\cal{L}}}
\newc{\LH}{\hat{L}}
\newc{\RH}{\hat{R}}
\begin{document}
\title{Unconventional Neutrino Mass Generation, Neutrinoless Double Beta Decays, and Collider Phenomenology}
\author{Chian-Shu Chen$^{1,2}$,
C.Q. Geng$^{1,2}$ and J. N. Ng$^{2}$}
\affiliation{
$^{1}$
Department of Physics, National Tsing Hua University, Hsinchu, Taiwan 300\\
$^{2}$Theory group, TRIUMF, 4004 Wesbrook Mall, Vancouver, B.C. V6T 2A3, Canada}
\date{\today}
\begin{abstract}
We study a model, in which lepton number violation is solely triggered  by a dimension 4 hard breaking term
in the scalar potential. A  minimal model which contains a $SU(2)$ triplet with hypercharge $Y=2$, and
a  pair of singlet doubly charged scalar fields in addition to the Standard Model (SM) Higgs doublet is 
constructed.  The model is
technically natural in the sense that lepton number is preserved in the limit that the hard term vanishes.
SM phenomenology restricts the vacuum expectation value of the triplet scalar field $v_T<5.78$ GeV.
Neutrino masses controlled by $v_T$ are  generated at the two loop level and are naturally 
to  the sub-eV range. In general they exhibit  normal hierarchy structure. Here the neutrino mass term does not
dominate  neutrinoless
double beta decays of nuclei. Instead the short distance physics with doubly charged Higgs exchange gives the
leading contribution.
We expect weak scale singly and doubly charged Higgs bosons to make their appearances at the LHC
and the ILC.

\end{abstract}

\maketitle

It is now generally accepted that the three active left-handed neutrinos of the Standard Model have very small
masses. However, the origin of this is still unknown. The orthodox view is they arise from 
the seesaw mechanism employing one or more very massive right-handed neutrinos ($> 10^{12}$ GeV). 
While this can be elegantly incorporated in some Grand Unified Theories such as $SO(10)$ direct
experimental tests of the mechanism is difficult if not impossible. There are attempts
to lower the seesaw mass scale to the TeV range, but this usually entails 
 complicated structures
in the right-handed neutrinos with fine tuning embedded. 

In this paper we investigate an alternative approach in which no right-handed neutrinos are involved.
We follow the proposal first investigated in \cite{Zee} that active neutrino masses arise from an extended Higgs sector of the SM without extending the gauge group. In this prototype construction a $SU(2)$ singlet Higgs field
with non-trivial hypercharge was used.  Then  doubly charged scalars were added later \cite{More} to give more
realistic neutrino masses. Similarly, we extend the Higgs sector by adding one $SU(2)$ singlet and one
triplet fields. We also postulate that the only lepton number violating interaction
are of dimension four and resides in the extended Higgs potential. Together with additional Yukawa terms
that are
allowed by the symmetry of the model we can generate light neutrino masses without adding singlet fermions.
 Since 
there are no right-handed
neutrinos in our construction, active neutrino masses must necessarily be of Majorana type.

Next we discuss the construction of our minimal model.  Group theory
dictates that $SU(2)$ singlets and/or triplets be involved. Indeed we introduce a triplet $T(1,2)$ and a 
singlet $\Psi(0,4)$, where the bracket denotes  $SU(2)\times U(1)$ quantum numbers. Then $T$ consists
the fields $T^0,T^{\pm},T^{\pm\pm}$ and $\Psi$ are doubly charged scalar fields $\Psi^{\pm\pm}$. We also
assign lepton number $L=2$ to $\Psi$ whereas $T$ fields carry no lepton number. With these ingredients
the most general renormalizable potential is given by
\begin{eqnarray}
\label{HLVV}
V(\phi,T,\psi) &=& -\mu^2\phi^{\dag}\phi+\lambda_{\phi}(\phi^{\dag}\phi)^2-\mu^2_TTr{(T^{\dag}T)}\nonr \\
&&+\lambda_T[Tr(T^{\dag}T)]^2+\lambda'_TTr(T^{\dag}TT^{\dag}T) \nonr \\
&& +m^2\Psi^{\dag}\Psi+
\lambda_{\Psi}(\Psi^{\dag}\Psi)^2 +\kappa_1Tr(\phi^{\dag}\phi T^{\dag}T)\nonr \\
&&+\kappa_{2}\phi^{\dag}TT^{\dag}\phi+\kappa_{\Psi}\phi^{\dag}\phi\Psi^{\dag}\Psi \nonumber \\
&& +\rho Tr(T^{\dag}T\Psi^{\dag}\Psi)
\nonumber\\
&&+
[ \lambda(\phi^{T}T\phi\Psi^{\dag})-M(\phi^{T}T^{\dag}\phi)+h.c.]\,,
\end{eqnarray}
where $\phi$ denotes the SM Higgs doublet field. The lepton number violation is given by the last term in Eq. (\ref{HLVV}).
In order to induce spontaneous symmetry breaking we take $\mu^2$, $\mu_{T}^2$ both
positive. Minimizing the potential gives us the vacuum expectation values: $\langle \phi^0 \rangle \equiv \frac{v}{\sqrt 2}$ and $\langle T^0 \rangle \equiv \frac{v_T}{\sqrt 2}$.
 
After SSB the $W$ and $Z$ bosons pick up masses at the tree level given by 
$M_W^2= \frac{g^2}{4}(v^2+2v_T^2)$ and $M_Z^2=\frac{g^2}{4\cos^2\theta_{W}}(v^2+4v_T^2)$ \cite{ChengLi}
where we have used standard notations. The tree level relation $e=g\sin \theta_W$ also holds. To get
an estimate on $v_T/v$ we use the limit from the Particle Data Group $\rho=1.002^{+.0007}_{-.0009}$
\cite{PDG} and the W mass from the Fermi coupling. 
 This gives approximately $v_T<5.78 \mathrm {GeV}$. As we shall see next this is 
a controlling scale for neutrino masses. As can be seen later, the dimension three soft term also does not
contribute to neutrino masses and by assuming no fortuitous cancellations in the minimal conditions
we obtain $M \gtrsim 7.5$ TeV.

It is also clear from Eq.(\ref{HLVV}) that there are mixing among various Higgs fields. In particular
$\mathfrak{Re} \phi^0, {\mathfrak{Re}} T^0$ pair will mix to give two physical neutral scalars $h^0, P^0$. The pair $\mathfrak{Im} \phi^0,
\mathfrak{Im} T^0$ will mix with one combination eaten by the $Z$ boson leaving a physical psuedoscalar $T^0_a$.
Similarly, for the charged states $\phi^{\pm}, T^{\pm}$ one combination will be  eaten by the $W$ bosons
leaving only  a pair of singly charged $P^{\pm}$ scalars. Finally the weak eigenstates $T^{\pm\pm}, \Psi^{\pm\pm}$ 
will also mix to form physical states $P_1^{\pm\pm},P_2^{\pm\pm}$ of masses denoted by $M_1 , M_2$
respectively and mixing angle $\omega$. All the masses and mixing angles are free parameters 
and we can use them to replace the parameters in $V(\phi,T,\Psi)$. They are to be determined experimentally.

Besides the usual Yukawa interactions of $\phi$ with the fermions we also 
add the term $Y_{ab}\overline{l^c_{aR}}l_{bR}\Psi$, which is lepton number conserving in our scheme and 
is allowed. Here $a,b$ are family indices. 
On the other hand, the term $LLT$ where $L$ denotes a SM lepton doublet violates lepton number and is disallowed
\cite{LLT}.
 The absence of this term evades the necessity 
of having to put in by hand  a very small value of $v_T$ in the eV range that  plagues other Higgs models of neutrino masses.
 Moreover, this term will be generated radiatively after symmetry breaking and is  small.
 Adding in  the SM terms and the covariant derivatives of
$T$ and $\Psi$ we have a renormalizable model.

It is well known that the Yukawa couplings of $\phi$ to fermions are diagonalized by a biunitary
transformation $U_L$ and $U_R$ so that the charged leptons
are mass eigenstates. Clearly, applying these rotations in general does not diagonalize $Y_{ab}$. Hence, we
expect flavor violating couplings $ Y'_{ab}$ between families of right-handed leptons and the physical $P^{++}$
states. Thus, the decay modes such as $P^{++}\ra \mu^+ e^+$ must occur in general. On the other hand, 
$P^{\pm}$ coupling to fermions is similar to SM but is scaled by $ (v_T/v)$.

\begin{figure}[ht]
  \centering
    \includegraphics[width=0.4\textwidth]{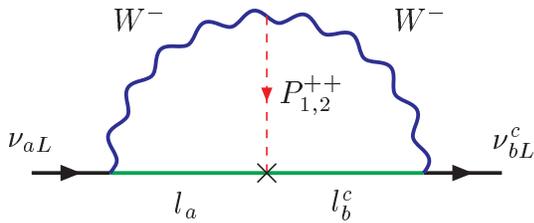}
  \caption{2-loop diagram for neutrino mass. }
  \label{fig:2loop}
\end{figure}

Now we can calculate the active neutrino mass matrix. The leading contribution is given by the 2-loop Feynman diagram 
depicted in Fig.(\ref{fig:2loop}).
We find
\beqa
(m_\nu)_{ab}
 \simeq \sqrt 2 g^4  m_a m_b v_T Y_{ab}\cos \omega \sin \omega 
 \ \ \ \ \ \ \ \ \ \ \ \ \ \ \ \ \ \ \ \ \ 
 \nonr \\
 \times \left[ I(M_W^2,M_1^2,m_{a},m_{b})- I(M_W^2,M_2^2,m_{a},m_{b})\right]\ \ \ 
\label{numass}
\eeqa
where
$a,b=e,\mu,\tau$.
The integral I is given by
\begin{eqnarray}
I(M_{W}^2,M_{i}^2,m_a^2,m_b^2)=
 \ \ \ \ \ \ \ \ \ \ \ \ \ \ \ \ \ \ \ \ \ \ \ \
  \ \ \ \ \ \ \ \ \ \ \
\nonr\\
\int\frac{d^4q}{(2\pi)^4}\int\frac{d^4k}{(2\pi)^4}\frac{1}{k^2-m^2_{a}}\frac{1}{k^2-M_{W}^2} \frac{1}{q^2-M_{W}^2}\ \ \ \ \nonr \\
\times \frac{1}{q^2-m^2_{b}}\frac{1}{(k-q)^2-M_{i}^2}\,. \ \ \ \ \ \  
\end{eqnarray}
There is a generalized G.I.M. mechanism at work here. This can clearly be seen 
in the limit $M_{1,2}>M_{W}$ \cite{2loop}:
\begin{equation}
 I(M_{W}^2,M_{i}^2,0,0) \sim \frac{1}{(4\pi)^4}\frac{1}{M_{i}^2}\ln^2 \left(\frac{M_{W}^2}{M_{i}^2}\right).
\label{intg}
\end{equation}
It is interesting to note that the $m_\nu$ mass matrix is independent of the singly charged Higgs 
at the leading order.
The lepton violating parameter is now hidden in the mixing angle $\omega$. In the limit $\lambda \ra 0$
then $\omega \ra 0$ and lepton number is restored and neutrinos remain massless. This can also
be seen by expressing Fig.(\ref{fig:2loop}) in weak eigenstates. The neutrino mass matrix is now proportional to the form 
\begin{equation}
\begin{pmatrix}
x_e^2Y_{ee}& x_e x_\mu Y_{e\mu}& x_e x_\tau Y_{e\tau} \\
x_e x_\mu Y_{e\mu}& x_{\mu}^2Y_{\mu\mu} & x_\mu x_\tau Y_{\mu\tau} \\
x_e x_\tau Y_{e\tau} & x_{\mu} x_{\tau} Y_{\mu\tau} & x_{\tau}^2  Y_{\tau\tau} 
\end{pmatrix} ,
\label{Mnu}
\end{equation}
where $x_a= m_a^2/M_W^2$ for $a=e,\mu, \tau$.
Barring fine tuning of $Y_{ab}$ this is of the normal hierarchy type since the first row and column
is much smaller than the rest due to the electron mass.
Before we embark on an estimate
of the size of the neutrino mass we list the experimental limits on the couplings $Y_{ab}$ \cite{Ylim}.
For flavor diagonal terms the most stringent ones are for the muon and electron sectors
\beqa
Y_{ee}^2
&\lesssim & 9.7 \times 10^{-6} {\mathrm {GeV}}^{-2} M^2_{P^{--}} \,,\nonr \\
Y_{ee}Y_{\mu\mu}
&\lesssim & 2.0\times 10^{-7}{\mathrm {GeV}}^{-2} M^2_{P^{--}}\,, \nonr \\
Y^2_{\mu\mu}
 &\lesssim & 2.5 \times 10^{-5} {\mathrm {Gev}}^{-2} M^2_{P^{--}}\,,
\label{Ydiag}
\eeqa
where $M^{-2}_{P^{--}}=\sin^{2}\omega M^{-2}_{1}+\cos^{2}\omega M^{-2}_{2}$ and we have updated 
the limits as given in \cite{Ylim2}.  
The corresponding limits for $\tau$ couplings are much less stringent.
These  limits are to  the masses  of the doubly charged scalars and their mixings. If we assume
that the lighter  one is  a 500 GeV particle and the mixing is small, then we get $Y_{ee}\lesssim 1.6$.

For non-diagonal terms the limits are more stringent from $\mu\ra 3e$
and $\mu\ra e\gamma$ decays \cite{Ylim2}. Explicitly 
\beqa
Y_{e\mu}Y_{ee}
&\lesssim &  6.6 \times 10^{-11} {\mathrm {GeV}}^{-2} M^2_{P^{--}}\,,
\nonr \\
Y_{l\mu}Y_{l\mu}
&\lesssim &  1.5 \times 10^{-9} {\mathrm {GeV}}^{-2} M^2_{P^{--}}\,,
\eeqa
where the second one is obtained from $\mu\ra e \gamma$, which is not as  tight since 
it is a 1-loop process \cite{mtoeg}.
Similar constraints on the $\tau$ are much weaker and  are given below:
\beqa
Y_{e\tau}Y_{ee}&<&3.0\times 10^{-8}\,GeV^{-2}M_{P^{--}}^{2}\,, \nonr \\
Y_{e\tau}Y_{\mu\mu}&<&3.0\times 10^{-8}\,GeV^{-2}M_{P^{--}}^{2}\,, \nonr \\
Y_{\tau\mu}Y_{\mu\mu}&<& 2.9\times 10^{-8}\,GeV^{-2}M_{P^{--}}^{2}\,,\nonr \\
Y_{\tau\mu}Y_{ee}&<&2.9\times 10^{-8}\,GeV^{-2}M_{P^{--}}^{2}\,.
\label{taulim}
\eeqa
We can now estimate the size of the $\tau \tau$
element of $m_\nu$. Explicitly,
\begin{equation}
 (m_\nu)_{\tau\tau} \sim 0.55 \left(\frac{Y_{\tau\tau}}{1}\right)\left(\frac{\sin
\omega}{0.1}\right)\left(\frac{v_T}{6 \mathrm{GeV}}\right) \mathrm {eV}
\end{equation}
where we used $M_1=0.5, M_2= 1 \mathrm {TeV}$. This is well within the experimental bounds from tritium
beta decay \cite{Kraus}. We recapitulate the physics for the smallness of $m_\nu$. It arises firstly from
a low triplet scale as required by  phenomenology, secondly from the 2-loop factor and thirdly
from the helicity flips in the internal charged lepton, which introduce two lepton masses, both of which are small.
Clearly, the $ee$ element will be much smaller due to the electron mass
suppression. Thus, the effect from $(m_\nu)_{ee}$ will not be detectable in the next round of
neutrinoless double beta decay, $0\nu\beta\beta$, of nuclei experiments.

\begin{figure}[ht]
  \centering
    \includegraphics[width=0.3\textwidth]{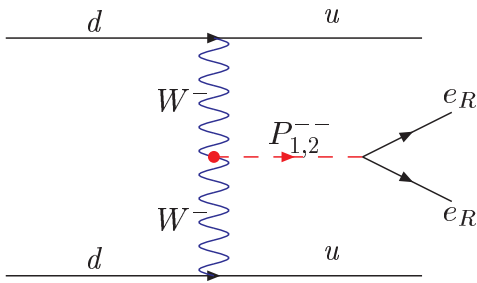}
  \caption{ Feynman diagram for Neutrinoless double beta decay }
  \label{fig:0nu}
\end{figure}

Next we examine the short distance physics itself in $0\nu\beta\beta$. The tree level Feynman diagram is
depicted in Fig.(\ref{fig:0nu})
. The amplitude is given by
\begin{equation}
A_{tree}\sim \frac{g^4Y_{ee}}{8\sqrt{2}}v_T\sin{\omega}\cos{\omega}\frac{1}{M^4_W}(\frac{1}{M^2_1}-\frac{1}{M^2_2})
\end{equation}
On the other hand, the amplitude from neutrino mass is given by \cite{0nup}
\begin{equation}
A_{\nu} \sim \frac{g^4}{16M_W^4}\frac{m_{\nu,ee}}{\langle p^{2} \rangle}
\end{equation}
where $\langle p^{2} \rangle \sim 0.01$ GeV$^{2}$. From Eqs.(\ref{numass},\ref{intg}) we can see clearly that
$A_{tree}>>A_{\nu}$. This can also be seen diagramatically. Fig.(\ref{fig:0nu}) can be obtained from
Fig.(\ref{fig:2loop}) by cutting the internal lepton line and attaching the $W$ propagators to quarks.
In doing so, the loop factors and the helicity flips, which are essential for 
$m_\nu$, no longer 
apply. Thus, the direct short distance physics is more important here, although the origins for
$0\nu\beta\beta$ and $m_\nu$ come from the same source in  the Higgs potential. 

To understand the connection between neutrino mass and $0\nu\beta\beta$ decays better we
examine a representative model in which TeV scale right-handed Majorana neutrinos are invoked to
generated small masses to the active neutrinos via the seesaw mechanism. In this case it is found that 
the exchange of active neutrino contribution dominates the $0\nu\beta\beta$ amplitudes \cite{TeVN}. 
Notice now  $(m_\nu)_{ab}$ is generated at the tree level.  A second example is given in \cite{SV,M-W} where a triple
Higgs similar to our $T$ is introduced together with lepton violation in $LLT$ term. In this model
$m_\nu$ is generated at tree level and the effective $W-W$-doubly charged Higgs coupling is proportional
to $m_\nu$ allowing the neutrino mass exchange to dominate in $0\nu\beta\beta$ process.
Our model gives an example, in which the short distance physics is now the leading term for $0\nu\beta\beta$ 
decays, since it is a tree level process. On the other hand, neutrino masses are dynamically generated
in higher loops as well as Yukawa suppressed. Thus, we conclude that in general
one cannot extract the value of the neutrino mass from the observation of $0\nu\beta\beta$ decays
aside from nuclear physics uncertainties.  However, a positive signal from these reactions clearly
indicates that new physics of  total lepton number violation is at work.
To determine the source, be it heavy Majorana neutrinos
or lepton violation in the scalar sector  as in our model or other models, 
will require further experimentation. 
Indeed, current limit on this decay will set $Y_{ee}<0.25$  for $M_1= 500$ GeV and the parameters
we used above. 

One of the most spectacular signal of the model we presented is the production of 
the doubly charged scalars $P^{\pm\pm}_{1,2}$. At the LHC they can be produced via the
2W fusion process:
\begin{equation}
u+u\ra d + d + P^{++} \ra d+d+ \tau^+ \mu^+
\label{UP} 
\end{equation}
and 
\begin{equation}
\label{DP}
d+d\ra u + u + P^{--} \ra u+u+ \tau^- \mu^-\,.
\end{equation}
The mechanism is the same as that depicted in Fig.(\ref{fig:0nu}) with obvious changes.
The signature for both is a pair of jets plus two resonating leptons. All six combinations
of lepton pair flavors are possible if no $Y_{ab}$ vanishes.  The Drell-Yan production mechanism is
much smaller here, since $P^{\pm\pm}$ do not directly couple to quarks. 
We expect the rate for Eq.(\ref{UP}) to twice  that of Eq.(\ref{DP}) due to the larger
u-quark content in proton. The lepton number violation in the final state will be unmistakable.
If the polarizations of the leptons can be measured, they can be used to distinguish the $P$
from other models with doubly charged Higgs bosons. Another characteristics of the model is
that $P_{1,2}$ will have small branching ratios into quark pairs. The expected production cross section is
$\sim 1.5$ fb for $M_1= 500$ GeV and the discovery limit at the LHC is $\sim 1.75 $ TeV \cite{Huitu}.

At the ILC it will be best to employ  the $e^- e^-$ option. The $P^{\pm\pm}$'s can be produced at rest and decay
into same sign lepton pairs with various flavor combinations. The signatures are clean and
will give a direct measurement of the couplings $Y_{ab}$. Whether they are observable  depends
crucially on $Y_{ee}$ as well as energetics. 
If the only option at the ILC is  $e^+e^-$ then the production of $P^{--}$ can proceed via
\begin{equation}
e^+e^- \ra e^+ l^+ +P^{--}
\end{equation}
with the subsequent decay of $P^{--}$ into lepton number violating channels. The signature is equally spectacular.
We emphasize that the   productions of $P^{\pm\pm}$ at the LHC and ILC
 will be direct checks on
the mechanism for $0\nu\beta\beta$ decays of nuclei, if they are observed first. They are crucial for
understanding the origin of neutrino masses.

Although the singly charged scalars $P^{\pm}$ play only a sub-dominant role in neutrino masses, they are
an essential part of the model. The important feature that distinguish $P^+$ from other charged Higgs
model is the universal reduction of its  Yukawa coupling to fermions. In particular the $t-b-P$ coupling is
given by 
\begin{equation}
\frac{\sqrt 2 v_T}{v^2} m_t \bar{b}_Lt_RP^{-} +h.c.
\end{equation}
This leads to its production at the LHC being dominated by the gluon b-quark fusion process:
\begin{equation}
b+g\ra t+P^-\,.
\end{equation}
The production rate will be analogous to that of two Higgs doublet models (2HDM)
with large $\tan\beta\sim 40$
\cite{Kidon} with 
 appropriate modifications to couplings. For a 400 GeV $P^{+}$ we expect the production cross
section at the LHC to be $\sim 0.1$ pb. The decay signature will depend on the mass. If it
is heavy enough  to decay into a t-b pair then this will be the dominant mode. Otherwise; the
main fermion decay mode will be into $\tau \nu$. Interestingly the helicity of the final state $\tau$ will be
exclusively right-handed. This can be used as a diagnostic tool for distinguishing $P^{\pm}$ from charged Higgs
in the 2HDM.

We have constructed a model of hard lepton number violation in the scalar sector by extending the SM Higgs
sector with a Higgs triplet and a singlet with two units of charged. We make essential use of the
coupling term $\Psi e_R e_R$ to generate active neutrino masses at the 2-loop level, and at the
same time $0\nu\beta\beta$ decays of nuclei are induced at tree level. This construction 
gives an example that these latter decays probe the short-distance physics of doubly charged Higgs
exchange and not  the exchange of light active neutrinos. We also sketched the phenomenology
of this extended Higgs model at the colliders and found it to be different from that of the 2HDM.
LHC can play a crucial role in understanding  the origin of neutrino masses if their governing scale
is not the GUT scale. It is also clear that flavor violating decays of charged leptons
should  be pushed further.

In conclusion, we note that there is another model of hard lepton number violation that does not
use doubly charged Higgs bosons. This model consists of a triplet $t^+,t^0,t^-$ with $Y=0$ and a 
singlet $S^{\pm}$ with $Y=2$. Assigning $L=0$ to the triplet and $L=2$ to the singlet allows the term
$LLS$ and the lepton number violation being triggered by $\phi^T t \phi S$. This resembles the
original Zee model for neutrino masses \cite{Zee}. Detail study of this second model will be left for a 
future study.\\

We would like to thank Prof D. Frekers for reading the manuscript and the many discussions we have
on the connection between neutrino masses and $0\nu\beta\beta$ decays that partly motivated this 
study. This work is supported in part by 
the Natural Science and Engineering Council of Canada
and
the National Science Council of
R.O.C. under Grant \#s:
NSC-94-2112-M-007-(004,005) and NSC-95-2112-M-007-059-MY3.


\begin{thebibliography}{99}
\bibitem{LLT}
There are several ways to naturally suppress
the Yukawa couplings of $L_{a}L_{b}T$. In extra dimensional models with split fermions
one can arrange the doublet to be separated more then the singlet, while leaving the Higgs
field on the TeV brane \cite{ExtraD}. Alternatively, one can arrange the triplet Higgs profile
to have smaller overlap with $L$ then with $l_R$ \cite{Tait}. One can
also extend our Higgs sector to include two Higgs doublets  and 
employ discrete symmetry to forbid
this term.
\end{thebibliography}
\end{document}